# Strong Superconducting Proximity Effect in Pb-Bi$_2$Te$_3$ Hybrid Structures

Fanming Qu*, Fan Yang*, Jie Shen*, Yue Ding, Jun Chen, Zhongqing Ji, Guangtong Liu, Jie Fan, Xiunian Jing, Changli Yang and Li Lu†

*Daniel Chee Tsui Laboratory, Beijing National Laboratory for Condensed Matter Physics & Institute of Physics, Chinese Academy of Sciences, Beijing 100190, People's Republic of China*

\* These authors contributed equally to this work.
† Corresponding authors: lilu@iphy.ac.cn

**Majarona fermions (MFs) were predicted more than seven decades ago but are yet to be identified [1]. Recently, much attention has been paid to search for MFs in condensed matter systems [2-10]. One of the seaching schemes is to create MF at the interface between an *s*-wave superconductor (SC) and a 3D topological insulator (TI) [11-13]. Experimentally, progresses have been achieved in the observations of a proximity-effect-induced supercurrent [14-16], a perfect Andreev reflection [17] and a conductance peak at the Fermi level [18]. However, further characterizations are still needed to clarify the nature of the SC-TI interface. In this Letter, we report on a strong proximity effect in Pb-Bi$_2$Te$_3$ hybrid structures, based on which Josephson junctions and superconducting quantum interference devices (SQUIDs) can be constructed. Josephson devices of this type would provide a test-bed for exploring novel phenomena such as MFs in the future.**

Bi$_2$Te$_3$ single crystals were grown by Bridgeman method, and were confirmed to be of high quality by X-ray diffraction (Fig. 1a). The carriers are of electron type, with a concentration of ~2×10$^{18}$ cm$^{-3}$ as determined from Hall effect and Shubnikov–de Haas (SdH) oscillation measurements at 2 K (Fig. 1b). Thin flakes of Bi$_2$Te$_3$ with typical sizes of 10 μm in length/width and 100~300 nm in thickness were exfoliated from a bulk crystal and transferred onto Si/SiO$_2$ substrates. Pb electrodes of ~150 nm thick were fabricated onto the flakes via standard e-beam lithography, magnetron sputtering and lift-off techniques. To study the proximity effect along the lateral direction of the flakes, parallel Pb electrodes were directly deposited on top of the flakes (see insets of Figs. 2a and 2b). A more complicated sandwich structure (see Figs. 3a and 3b) was employed to study the penetration of proximity effect in the thickness direction. The measurements were performed at low temperatures down to 15 mK in a dilution refrigerator or 250 mK in a $^3$He cryostat. The differential resistance between two selected electrodes was measured as a function of both dc bias current and applied magnetic field with a quasi-four-probe measurement configuration. Precise control of magnetic field to milli-Gauss level is achieved by using a Keithley 2400 source meter to drive the superconducting magnet of the cryostat.

To study the superconducting proximity effect along lateral direction, we have fabricated and investigated more than a dozen Josephson junctions that have parallel Pb electrodes of various lengths and inter-electrode distances deposited onto the top surface of the flakes. The insets of Figs. 2a and 2b show the scanning electron microscope (SEM) images of two types of junctions, one with Pb electrodes across the whole width of the flakes (junction #1, categorized as type-I), and the other with Pb electrodes located in the middle of the upper surface, away from the edges (junction #2, categorized as type-II).

Figures 2a and 2b show respectively the measured differential resistance *dV/dI* of the two junctions as functions of both magnetic field *B* and bias current $I_{dc}$. The dark blue region represents the zero-resistance superconducting state. It can be seen that the critical current $I_c$ is modulated by magnetic field, demonstrating a Fraunhofer-like pattern as expected for an *s*-wave type Josephson junction [19]:

$$I_c(B) = I_c(0) \left| \sin(\frac{\pi \Phi_J}{\phi_0}) / (\frac{\pi \Phi_J}{\phi_0}) \right|$$

where $I_c(0)$ is the critical current of the junction at zero magnetic field, $\Phi_J$ is the flux through the junction area, and $\phi_0 = h/2e$ is the flux quanta.

The measured oscillation periods of Fraunhofer pattern were consistent with the junction areas for all junctions. For example, the measured oscillation period for junction #2, $\Delta B$=1.45 G, corresponds to an effective junction area: $S_{eff} = \phi_0/\Delta B$ =14.3 μm$^2$, which is consistent with the actual junction area $S$=13.8 μm$^2$ estimated from the length (6 μm) and the center-to-center distance (2.3 μm) of the Pb electrodes (with the flux compressing taken into account).

For type-I junctions, the experimental data agree well with the theoretical Fraunhofer pattern (see, e.g., Fig. 2a for junction #1), with the ratio of the second peak height to the main peak height close to the expected value of 0.22. For type-II junctions, however, the measured oscillation patterns deviate significantly from the expectation, with the peak height ratio ranging from 0.08 to 0.11 for seven such junctions investigated (the ratio is 0.1 for junction #2, see Fig. 2b). The results indicate a non-uniform spatial distribution of the supercurrent in type-II junctions [20], presumably due to stray supercurrent flowing through the area near the two sides of the junctions.

Figure 2c shows one of the current vs. voltage ($I_{dc}$ vs. $V$) curves of junction #2 at 15 mK. A hysteretic loop is seen in bi-directional current sweepings. We note that the asymmetric $I_{dc}$-$V$ curve in uni-directional current sweeping is not likely to be caused by self-heating, because hysteretic behavior remains in junctions with a very small critical current (for example, junctions with a 1.5 μA critical current still exhibits an obvious hysteretic loop, see Fig. S1 of Supplementary information). Instead, the hysteretic behavior would indicate the formation of an underdamped Josephson junction [21] between adjacent Pb electrodes. An underdamped behavior is usually seen in planar tunneling junctions with significantly large junction capacitance and shunted resistance [19, 21]. In our junctions, however, the capacitance should be negligibly small compared to planar tunneling junctions, and the junction resistance is only around one or several Ohms. Therefore, the appearance of a hysteretic behavior is quite unusual. It might be related to the strong spin-orbit coupling in $Bi_2Te_3$, being less dissipative because of reduced back scattering.

To examine how far a supercurrent can be established through proximity effect along the lateral direction, we have fabricated a number of junctions with different inter-electrode distances. For junction #2, which has an inter-electrode distance of 0.7 μm, a supercurrent held up to at least 1.5 K at zero magnetic field, as shown in Fig. 2d. For another junction with much longer inter-electrode distance of 3.5 μm (junction #3, SEM image not shown), a supercurrent held up to 120 mK, as shown in Fig. 2e. A 2D image plot of the differential resistance of junction #3 as a function of both magnetic field and bias current can be found in Fig. S2 of Supplementary information. The appearance of a proximity-effect-induced supercurrent over such a long distance would enable us to construct more complicated devices for testing various interesting physics.

To test whether a supercurrent can also be established via proximity effect along the thickness direction of the flakes, two sandwich-like devices were fabricated, and similar results were obtained. The device structure is shown in Fig. 3a and illustrated in Fig. 3b. A Pb film was firstly sputtered on one surface of the flakes when they were still on the Scotch tape after being exfoliated. Then those flakes of 100-300 nm in thickness were selected and transferred onto a $Si/SiO_2$ substrate, with the Pb side facing down to the substrate. Afterwards, an overexposed PMMA layer was coated to cover one side/edge of the selected flakes, serving as an insulation mask. Finally, two top Pb electrodes, with a same contacting area of ~3×2 μm², and separated by ~10 μm, were deposited on top of each flake. According to our previous investigation on lateral junctions, we know that no supercurrent can be established over ~10 μm distance along the lateral direction above 100 mK. Therefore, the measurement between the two top Pb electrodes actually probes the superconducting transition of the two sandwich junctions in the thickness direction.

Figure 3c shows the temperature dependence of quasi-four-probe resistance of device #4

between the two top electrodes. With decreasing temperature, the resistance of both devices undergoes two superconducting transitions around 7~8 K, one at a slightly higher temperature but with an obviously smaller resistance jump, the other at a lower temperature and with a larger resistance jump. These transitions are closely related to the superconducting transition of Pb, which should occur at $T_c$=7.2 K for pure bulk samples. The observed transition temperatures were slightly higher than that of pure Pb, indicating the possible formation of a rough or alloyed interface. Nevertheless, it is really a surprise that the bulk of $Bi_2Te_3$ flakes becomes completely superconducting at a temperature so close to the $T_c$ of Pb.

There are two possible scenarios for the occurrence of two transitions in the *R vs. T* curve in Fig. 3c. In the first scenario, the resistance jump at high temperature is caused by the superconducting transition of the bottom Pb film, which partially shorts out the flake, and the resistance jump at low temperature is due to the proximity-induced superconducting transition of the two sandwich junctions. In the second scenario, these two resistance jumps correspond to the individual superconducting transition of the two sandwich junctions, respectively.

The *dV/dI vs. $I_{dc}$* curve shown in Fig. 3d is in favor of the first scenario. As can be seen, two jumps with different critical currents appear on the *dV/dI vs. $I_{dc}$* curve. The two jumps there should all be ascribed to the superconducting transitions of the sandwich junctions, not that one to the superconducting transition of the junctions and one to the superconducting transition of the Pb film on the back, since the critical current of Pb film should be much larger. The data in Fig. 3d tell us that the resistances of the two sandwich junctions are about the same before becoming superconducting. We therefore believe that the two resistance jumps in the *R vs. T* curve, with different but well defined amplitudes because of their step-like shapes, are caused by different origins. We ascribe the jump occurred at higher temperature to the superconducting transition of the Pb film. Moreover, for device #5 (see Fig. S3 in Supplementary information) the lower transition in *R vs. T* curve actually contains double jumps, further supporting the first scenario.

To summarize, we have shown that a supercurrent can be easily established in the thickness direction of $Bi_2Te_3$ flakes via proximity effect, so that the whole volume in this direction needs to be considered when studying Josephson effect along the lateral direction of the flakes.

Based on the proximity-effect-induced superconductivity, more than ten SQUIDs with various shapes and areas were fabricated on the surface of $Bi_2Te_3$ flakes and measured down to dilution refrigerator temperatures. One of the typical results measured at 15 mK is shown in Fig. 4. The critical current of the SQUIDs shows standard interference patterns against magnetic field. In addition, the envelope of the interference pattern is modulated by the Fraunhofer diffraction patterns of each single junction. These behaviors can be described by the following formula [19]:

$$I_c(B) = 2I_c(0) \left| \sin(\frac{\pi \Phi_J}{\phi_0}) / (\frac{\pi \Phi_J}{\phi_0}) \right| \left| \cos \frac{\pi \Phi}{\phi_0} \right|$$

where $I_c(0)$ is the critical current of each single junction at zero magnetic field, $\Phi_J$ is the flux through the single junction area, and $\Phi$ is the flux through the ring area of the SQUID.

Both the periods of interference and Fraunhofer diffraction are found to be consistent with their corresponding areas, i.e., the areas of the ring and the junction, respectively. Specifically, from the data shown in Fig. 4, the observed periods are 0.48 G and 3.7 G, corresponding to areas of 43.2 $\mu m^2$ and 5.6 $\mu m^2$ that are in agreement with the measured areas of 38.5 $\mu m^2$ and 6.4 $\mu m^2$ for the ring and the junction, respectively. Since the effective areas of small SQUIDs are not easy to be determined if flux is being compressed by surrounding superconducting electrodes, a SQUID with a large area and thinner arms were investigated. The result is shown in Fig. S4 of Supplementary information. It confirms that the period-to-area relation for conventional SQUIDs holds accurately in Pb-$Bi_2Te_3$-Pb proximity-effect SQUIDs.

We found that both the interference pattern of the ring and the Fraunhofer diffraction pattern of the junctions are consistent with *s*-wave type of Josephson devices, i.e., the maximum of critical current is located at zero magnetic field. So far, no sign of unconventional pairing symmetry is recognized, presumably due to the dominating role of the Pb electrodes and/or the trivial superconducting feature of the bulk. More sophisticated device structures are demanded to reveal the possible unconventional pairing symmetry at the *s*-wave superconductor/topological insulator interface.

In summary, we have demonstrated that a Josephson-like supercurrent can be established over a distance of several microns in $Bi_2Te_3$ by Pb electrodes via strong proximity effect. We further demonstrate that SQUIDs can be conveniently constructed by utilizing this proximity effect. These Josephson devices exhibit an *s*-wave-like feature, presumably due to the dominating role of the Pb electrodes, and/or due to the trivial superconducting feature of the bulk. More sophisticated devices could be designed based on these results, to help searching for Majorana fermions in the future at a device level.

**Acknowledgments**

We would like to thank T. Xiang, L. Fu, G. H. Chen, Z. Fang and X. Dai for stimulative discussions. This work was supported by the National Basic Research Program of China from the MOST under the contract No. 2009CB929101 and 2011CB921702, by the Knowledge Innovation Project and the Instrument Developing Project of CAS, and by the NSFC under the contract No. 11174340 and 11174357.

**References**

1. Wilczek, F. Majorana returns. *Nature Phys.* **5**, 614-618 (2009).

2. Fu, L. & Kane, C. L. Superconducting proximity effect and Majorana fermions at the surface of a topological insulator. *Phys. Rev. Lett.* **100**, 096407 (2008).

3. Qi, X.-L., Hughes, T. L. & Zhang, S.-C. Chiral topological superconductor from the quantum Hall state. *Phys. Rev. B* **82**, 184516 (2010).

4. Sau, J. D., Lutchyn, R. M., Tewari, S. & Sarma, S. D. Generic new platform for topological quantum computation using semiconductor heterostructures. *Phys. Rev. Lett.* **104**, 040502 (2010).

5. Sato, M., Takahashi, Y. & Fujimoto, S. Non-Abelian topological order in *s*-wave superfluids of ultracold fermionic atoms. *Phys. Rev. Lett.* **103**, 020401 (2009).

6. Fu, L. & Kane, C. L. Probing neutral Majorana fermion edge modes with charge transport. *Phys. Rev. Lett.* **102**, 216403 (2009).

7. Akhmerov, A. R., Nilsson, J. & Beenakker, C. W. J. Electrically detected interferometry of

Majorana fermions in a topological insulator. *Phys. Rev. Lett.* **102**, 216404 (2009).

8.  Seradjeh, B., Moore, J. E. & Franz, M. Exciton condensation and charge fractionalization in a topological insulator film. *Phys. Rev. Lett.* **103**, 066402 (2009).

9.  Tanaka, Y., Yokoyama, T. & Nagaosa, N. Manipulation of the Majorana fermion, Andreev reflection, and Josephson current on topological insulators. *Phys. Rev. Lett.* **103**, 107002 (2009).

10. Law, K. T., Lee, P. A. & Ng, T. K. Majorana fermion induced resonant Andreev reflection. *Phys. Rev. Lett.* **103**, 237001 (2009).

11. Qi, X.-L. & Zhang, S.-C. The quantum spin Hall effect and topological insulators. *Phys. Today* **63**, 33-38 (2010).

12. Hasan, M. Z. & Kane, C. L. *Colloquium*: Topological insulators. *Rev. Mod. Phys.* **82**, 3045 (2010).

13. Moore, J. E. The birth of topological insulators. *Nature* **464**, 194-198 (2010).

14. Kasumov, A. Y. *et al.* Anomalous proximity effect in the Nb-BiSb-Nb junctions. *Phys. Rev. Lett.* **77**, 3029 (1996).

15. Sacépé, B. *et al.* Gate-tuned normal and superconducting transport at the surface of a topological insulator. Preprint at http://arxiv.org/abs/1101.2352 (2011).

16. Zhang, D. M. *et al.* Superconducting proximity effect and possible evidence for Pearl vortices in a candidate topological insulator. *Phys. Rev. B* **84**, 165120 (2011).

17. Knez, I., Du, R.-R. & Sullivan, G. Perfect Andreev reflection of helical edge modes in InAs/GaSb quantum wells. Preprint at http://arxiv.org/abs/1106.5819 (2011).

18. Yang, F. *et al.* Proximity effect at superconducting Sn-$Bi_2Se_3$ interface. Preprint at http://arxiv.org/abs/1105.0229 (2011).

19. Clarke, J. & Braginski, A. I. *The SQUID Handbook Vol. 1* (WILEY-VCH Verlag GmbH & Co. KGaA, Weinheim, 2004).

20. Dynes, R. C. & Fulton, T. A. Supercurrent density distribution in Josephson junctions. *Phys. Rev. B* **3**, 3015 (1971).

21. Heersche, H. B. *et al.* Bipolar supercurrent in graphene. *Nature* **446**, 56–59 (2007).

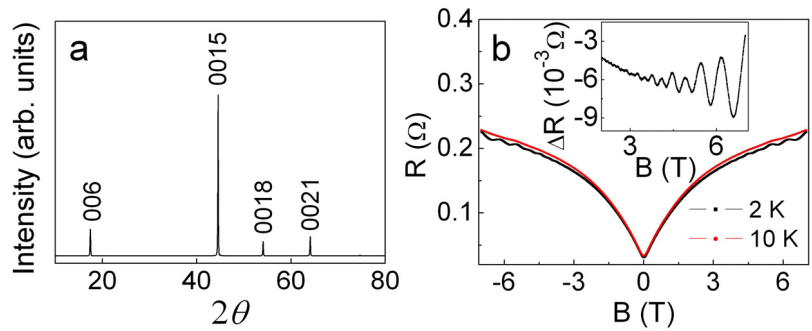

**Figure 1 | Characterizations of the Bi$_2$Te$_3$ crystal used in this experiment**. **a**, X-ray diffraction pattern of the crystal. **b**, Magneto-resistance of a Hall-bar shaped Bi$_2$Te$_3$ flake measured at 2 K (black) and 10 K (red) in perpendicular magnetic fields. The pronounced dip at zero magnetic field is caused by electron weak anti-localization due to strong spin-orbit coupling. Inset: difference of resistance between the 2 K curve and the 10 K curve, showing clearly the Shubnikov-de Haas oscillations at low temperatures, with an oscillation period of ~0.02 T$^{-1}$.

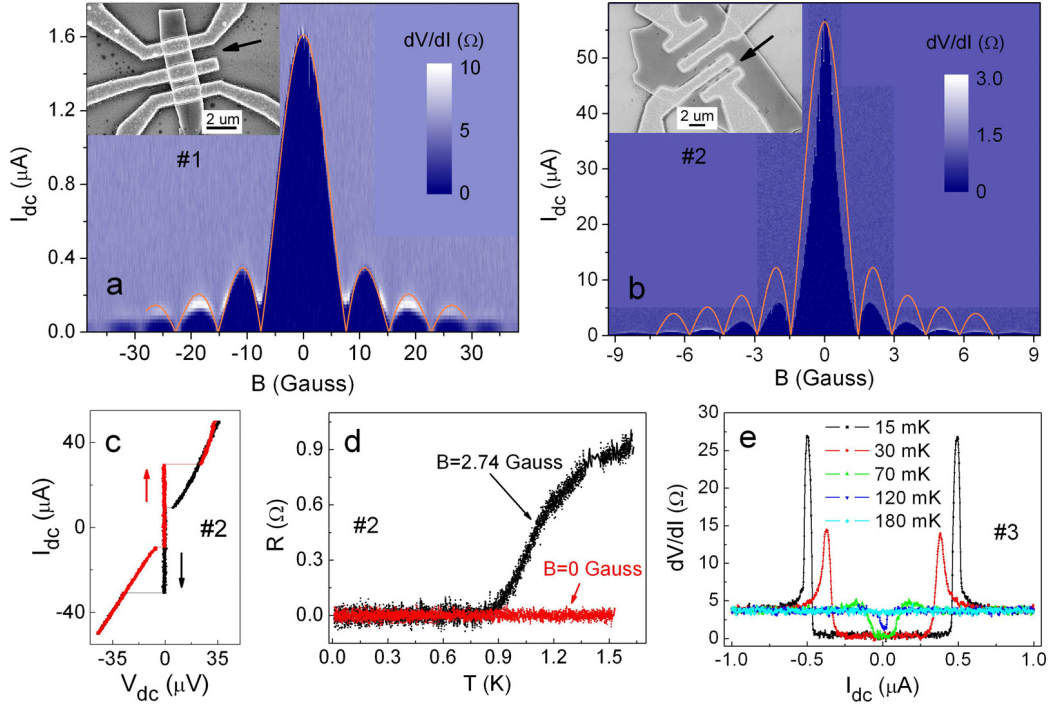

**Figure 2 | Proximity-effect-induced superconductivity in Pb-Bi$_2$Te$_3$-Pb structures along lateral direction. a**, Inset: SEM image of type-I lateral Josephson junctions, with parallel Pb electrodes across the whole width of the Bi$_2$Te$_3$ flake. Main frame: a 2D image plot of the differential resistance *dV/dI* of junction #1 (indicated by the arrow), measured at 15 mK as a function of both magnetic field and dc bias current. The boundary of *dV/dI* from dark blue to light blue, which separates the superconducting state from the finite-resistance normal state, oscillates with magnetic field following the theoretically expected Fraunhofer pattern (the orange curve). **b**, Inset: SEM image of type-II lateral Josephson junctions, with parallel Pb electrodes occupying the center part of the top surface. Main frame: The 2D image plot of *dV/dI* of junction #2 (indicated by the arrow), measured at 15 mK as a function of both magnetic field and dc bias current. The critical current oscillates with magnetic field in a way deviating from the standard Fraunhofer pattern. **c**, *I$_{dc}$ vs. V$_{dc}$* curve of junction #2 measured at 15 mK and in a magnetic field of 0.5 G, with bias current sweeping from positive to negative (black) and *vice versa* (red). Obvious hysteretic behavior can be seen. **d**, Temperature dependencies of the zero-bias resistance of junction #2, measured in two different magnetic fields. **e**, *dV/dI vs. I$_{dc}$* curves of junction #3 (type-I) composed of two Pb electrodes seperated by 3.5 μm, holding a supercurrent up to 120 mK.

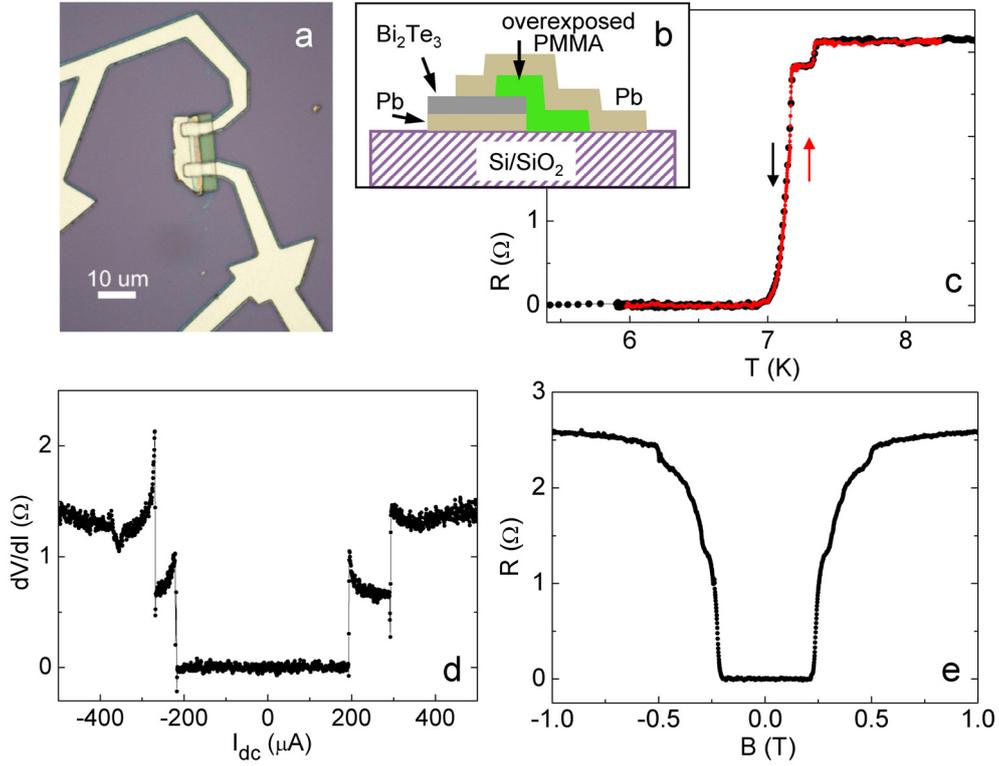

**Figure 3 | Proximity-effect-induced superconductivity in Pb-Bi$_2$Te$_3$-Pb sandwich structures. a**, Optical image of a sandwich-structured device (device #4). **b**, Illustration of the device structure. **c**, Temperature dependence of the zero-bias resistance of device #4 measured from one top Pb electrode to another with a quasi-four-probe measurement configuration. Both ramping down (black) and up (red) curves are shown. The resistance jump at higher temperature corresponds to the superconducting transition of the Pb film, and the jump at lower temperature corresponds to the superconducting transitions of the two sandwich junctions occurred simultaneously. **d**, A typical *dV/dI vs. I$_{dc}$* curve of device #4 measured at 0.3 K. The two jumps of *dV/dI* with similar amplitudes correspond respectively to the individual superconducting transition of the two sandwich junctions. **e**, The magnetic field dependence of the resistance of device #4 measured at 0.3 K and with *I$_{dc}$*=0.

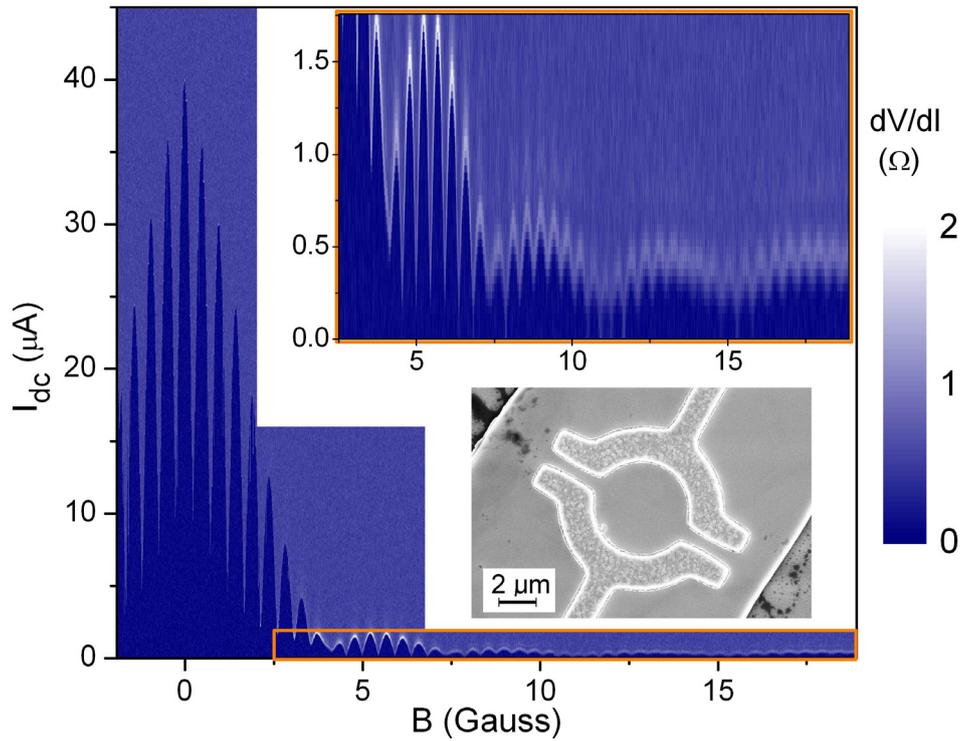

**Figure 4 | Superconducting quantum interference device (SQUID) based on the proximity-effect-induced superconductivity**. Lower inset: SEM image of a SQUID with Pb electrodes on the top surface of a $Bi_2Te_3$ flake. Main frame: differential resistance of the SQUID as a function of both magnetic field and bias current. Upper inset: Details of the orange rectangle region in the main frame, showing clearly the oscillations of critical current with magnetic field due to interference along the ring of the SQUID, and the modulation of the oscillations by the Fraunhofer diffraction pattern of the single junctions.

# Supplementary Information for "Strong Superconducting Proximity Effect in Pb-Bi$_2$Te$_3$ Hybrid Structures"

Fanming Qu*, Fan Yang*, Jie Shen*, Yue Ding, Jun Chen, Zhongqing Ji, Guangtong Liu, Jie Fan, Xiunian Jing, Changli Yang and Li Lu†

*Daniel Chee Tsui Laboratory, Beijing National Laboratory for Condensed Matter Physics & Institute of Physics, Chinese Academy of Sciences, Beijing 100190, People's Republic of China*

\* These authors contributed equally to this work.
† Corresponding authors: lilu@iphy.ac.cn

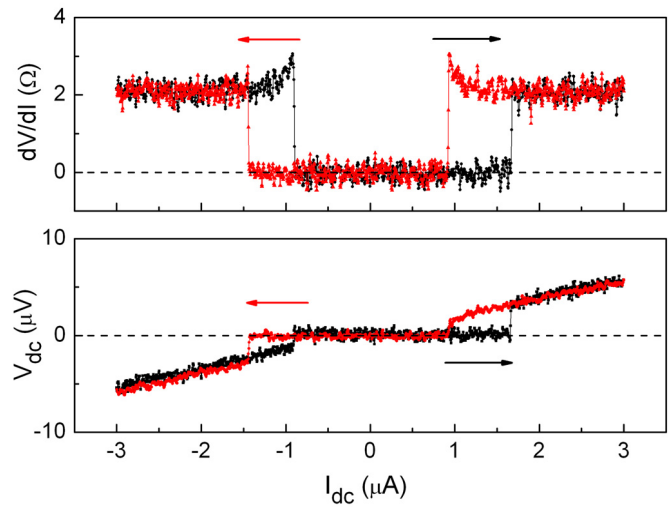

**Figure S1 | Hysteretic behavior of a Pb-Bi$_2$Te$_3$-Pb junction with a small critical current.** Differential resistance (*dV/dI*) vs. bias current (*I$_{dc}$*) and *V$_{dc}$* vs. *I$_{dc}$* curves with sweeping *I$_{dc}$* from negative to positive (black) and *vice versa* (red) for a lateral Pb-Bi$_2$Te$_3$-Pb junction measured at 15 mK. A hysteretic behavior occurs when the critical current is as small as 1.5 μA and with a normal state resistance of ~2 Ohm.

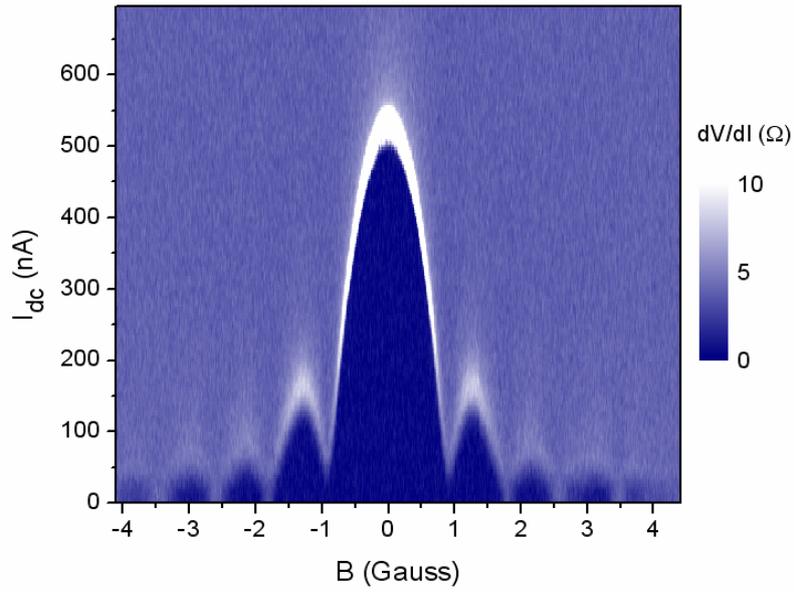

**Figure S2 | Fraunhofer pattern of a lateral Pb-Bi$_2$Te$_3$-Pb junction with two Pb electrodes seperated by 3.5 μm (junction #3 in main text).** The data were measured at 15 mK.

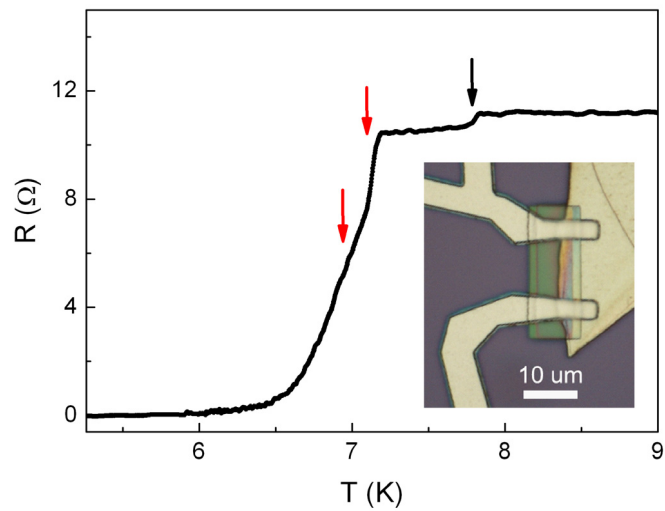

**Figure S3 | *R vs. T* curve of a Pb-Bi$_2$Te$_3$-Pb sandwich-structured device (device #5 in main text).** The three arrows indicate three resistance drops with decreasing temperature. The one at higher temperature (black arrow) corresponds to the superconducting transition of the Pb film. The other two drops (red arrows) correspond to individual superconducting transition of the two Pb-Bi$_2$Te$_3$-Pb sandwich junctions, respectively. The difference in transition temperature may come from the diffence in local environment of the two junctions. Inset: Optical image for the device.

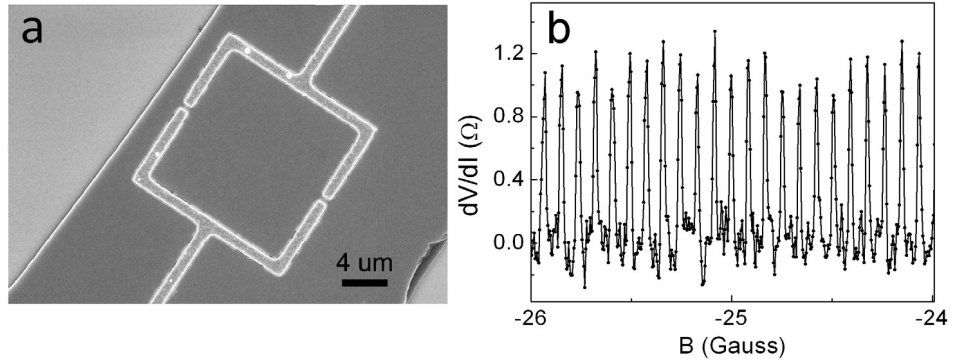

**Figure S4 | Investigation on the interference period of a large-area Pb-Bi$_2$Te$_3$-Pb SQUID. a,** Scanning electron microscope image of a 15 μm×15 μm SQUID, the largest one measured in this experiment. **b,** Differential resistance (*dV/dI*) as a function of magnetic field (*B*) of this device measured at 15 mK, with a biased current $I_{dc}$=200 nA. The oscillation period is $\Delta B$=0.085 G, corresponding to an effective area $h/(2e\Delta B)$=243 μm$^2$ which is consistent with the area of the device: *S*=225 μm$^2$.